# Nonextensibility of energy in Tsallis' statistics and the zeroth law of thermodynamics

Congjie Ou and Jincan Chen*

CCAST (World Laboratory), P. O. 8730, Beijing 100080, People's Republic of China and Department of Physics, Xiamen University, Xiamen 361005, People's Republic of China

Two important problems existing in Tsallis' statistics are investigated, where one is whether energy is extensive or not, and the other is whether it is necessary to introduce the so-called generalized zeroth law of thermodynamics or not. The results obtained show clearly that like entropy, energy is also nonextensive in Tsallis' statistics, and that the zeroth law of thermodynamics has been implicitly used in Tsallis' statistics since 1988. Moreover, it is expounded that the standard energy additivity rule adopted by a great number of researchers is not suitable in Tsallis' statistics, because it not only violates the law of energy conservation but also its corollary is in contradiction with the zeroth law of thermodynamics.

---

*Author to whom all correspondence should be addressed.

Mailing address: Department of Physics, Xiamen University, Xiamen 361005, People's Republic of China

Email: jcchen@xmu.edu.cn

Temperature, internal energy and entropy are three of the most important parameters in thermodynamics. The concepts of temperature and internal energy become nontrivial when entropy appears to be nonextensive. Since the generalized statistical entropy[1] was proposed by Tsallis in 1988, the non-extensibility of entropy in some complex systems with long-range interactions and/or long-duration memory has been widely recognized[2-8]. However, there are still two problems of physical importance in Tsallis' statistics, where one is whether the internal energies of these complex systems are extensive or not, and the other is whether it is necessary to introduce the generalized zeroth law of thermodynamics or not. Although the two problems have been discussed for many years[9-15], they have not been solved up to now, and consequently, have affected the development and improvement of Tsallis' statistics. Thus, it is very important and urgent to solve the two problems and reach some useful conclusions.

In nonextensive statistical mechanics developed from Tsallis' entropy[1], the first choice is very little used in the literature since it could not solve the relevant mathematical difficulties although there are three different choices for the internal energy constraint[16]. For the second choice[1-3, 8], $U_2 = \sum_i^w p_i^q \varepsilon_i$, the distribution function[17]

$$(p_2)_i = \frac{[1-(1-q)\beta\varepsilon_i]^{1/(1-q)}}{Z_2} \tag{1}$$

can be derived from the generalized statistical entropy[1]

$$S = -k(1-\sum_i^w p_i^q)/(1-q), \quad (\sum_i^w p_i = 1, \ q \in R) \tag{2}$$

where $Z_2 = \sum_i^w [1-(1-q)\beta\varepsilon_i]^{1/(1-q)}$, $k$ is a positive constant, $w$ is the total number of microscopic possibilities of the system, $\varepsilon_i$ is the energy of the system at the state $i$. Similarly, for the third choice[4,6,8], $U_3 = \sum_i^w p_i^q \varepsilon_i / \sum_i^w p_i^q$, one can obtain the distribution function as[16]

$$(p_3)_i = \frac{[1-(1-q)\beta(\varepsilon_i - U)/\sum_i^w (p_3)_i^q]^{1/(1-q)}}{Z_3} \tag{3}$$

with $Z_3 = \sum_i^w [1-(1-q)\beta(\varepsilon_i - U)/\sum_i^w p_i^q]^{1/(1-q)}$. It can be proved[2-4,6,8,16-18] that $\partial S/\partial U_l = k\beta = 1/T$, $(l=2,3)$, where $T$ is the absolute temperature. For the sake of convenience, $\sum_i^w$ is replaced by $\sum_i$ below.

For an isolated system C composed of two subsystems A and B of which the distributions satisfy[1,16-18]

$$p_{ij}(C) = p_i(A)p_j(B), \tag{4}$$

using the relations $\sum_i (p_2)_i^q = Z_2^{1-q} + (1-q)\beta U_2$ and $\sum_i (p_3)_i^q = Z_3^{1-q}$ and Eqs. (1)-(4), we can obtain the pseudo-additivity entropy rule [1,5,8,16]

$$S(C) = S(A) + S(B) + \frac{1-q}{k} S(A)S(B) \tag{5}$$

and the pseudo-additivity energy rules

$$\beta(C)U_2(C) = \beta(A)U_2(A)Z_2^{1-q}(B) + \beta(B)U_2(B)Z_2^{1-q}(A) + (1-q)\beta(A)\beta(B)U_2(A)U_2(B), \tag{6}$$

$$\beta(C)[\varepsilon_{ij}(C) - U_3(C)] = \beta(A)[\varepsilon_i(A) - U_3(A)]Z_3^{1-q}(B) + \beta(B)[\varepsilon_j(B) - U_3(B)]Z_3^{1-q}(A) \\ -(1-q)\beta(A)\beta(B)[\varepsilon_i(A) - U_3(A)][\varepsilon_j(B) - U_3(B)] \tag{7}$$

It is worthwhile pointing out that Eqs. (6) and (7) are two important results that have never appeared in Tsallis' statistics and one of the important bases for discussing and solving two problems mentioned in this paper.

When one important condition

$$\beta(C) = \beta(A) = \beta(B) \tag{8}$$

is adopted, Eqs. (6) and (7) may be, respectively, simplified as[9,18]

$$U_2(C) = U_2(A)Z_2^{1-q}(B) + U_2(B)Z_2^{1-q}(A) + (1-q)\beta U_2(A)U_2(B), \tag{9}$$

$$[\varepsilon_{ij}(C)-U_3(C)]=[\varepsilon_i(A)-U_3(A)]Z_3^{1-q}(B)+[\varepsilon_j(B)-U_3(B)]Z_3^{1-q}(A)$$
$$-(1-q)\beta[\varepsilon_i(A)-U_3(A)][\varepsilon_j(B)-U_3(B)] \quad (10)$$

It is important to note the fact that Eq. (8) has been implicitly used in nonextensive statistical mechanics[8-18] since the generalized statistical entropy[1] was advanced in 1988, although it has never been obviously given in literature of nonextensive statistical mechanics. One will find from the following analysis that Eq. (8) is essentially the mathematical expression of the zero law of thermodynamics.

Using the laws of entropy and energy conservation and the above equations, we can strictly prove

$$[1+\frac{1-q}{k}S(B)]\frac{\partial S(A)}{\partial U_l(A)}\delta U_l(A)+[1+\frac{1-q}{k}S(A)]\frac{\partial S(B)}{\partial U_l(B)}\delta U_l(B)=0, \quad (l=2,3) \quad (11)$$

and

$$\sum_j (p_l)_j^q(B)\delta U_l(A)+\sum_i (p_l)_i^q(A)\delta U_l(B)=0, \quad (l=2,3). \quad (12)$$

From Eqs. (11), (12) and (2), one obtains

$$\frac{\partial S(A)}{\partial U_l(A)}=\frac{\partial S(B)}{\partial U_l(B)}, \quad (l=2,3), \text{ or } \beta(A)=\beta(B). \quad (13)$$

It is just the zeroth law of thermodynamics. Obviously, the physical essence of Eq. (13) is completely identical with that of Eq. (8). It implies the fact that starting from Eq. (8), one get Eq. (13) which is the same result as Eq. (8). It is thus clear that the derivative process of Eq. (13) is only of a self-consistent calculation, but is not a proof for the zeroth law of thermodynamics in nonextensive statistical mechanics. This shows clearly that the zeroth law of thermodynamics still holds in nonextensive statistical mechanics, but it can not be proved from theory. Consequently, the concept of temperature is also suitable in nonextensive statistical mechanics.

It is also important to note the other fact that in nonextensive statistical mechanics, if Eq. (8) has not been implicitly used, one can not get the standard energy additivity rule[8, 10-16]

$$U(C) = U(A) + U(B), \tag{14}$$

even though the third term on the right hand side of Eqs. (6) and (7) is not considered. It is thus obvious that Eq. (8) is a necessary condition for the validity of Eq. (14). However, many researchers have not expounded the question and directly used Eqs. (1)-(5) and (14) to investigate some important problems. For example, they have been used to derive the so called generalized zeroth law of thermodynamics in nonextensive statistical mechanics. By comparing the expression of the so called generalized zeroth law of thermodynamics[8,11-16,18]

$$Z_3^{q-1}(A)\beta(A) = Z_3^{q-1}(B)\beta(B) \tag{15}$$

or

$$\frac{k\beta(A)}{1+[(1-q)/k]S(A)} = \frac{k\beta(B)}{1+[(1-q)/k]S(B)} \tag{16}$$

with Eq. (8), it can be seen without difficulty that Eq. (15) or (16) is obviously in contradiction with Eq. (8), because $Z_3^{q-1}(A)$ and $S(A)$ are not, in general, equal to $Z_3^{q-1}(B)$ and $S(B)$, respectively. This shows clearly that the standard energy additivity rule (14) which has been widely used by a lot of researchers may not be suitable in nonextensive statistical mechanics because its corollary violates the zeroth law of thermodynamics[19]. Therefore, it is unnecessary to introduce the so called generalized zeroth law of thermodynamics in nonextensive statistical mechanics and the new concept of the physical temperature[8,11-16,20].

The above results show clearly that, Eqs. (9) and (10) can be derived in nonextensive statistical mechanics，based on Eqs. (1)-(4) and the zeroth law of thermodynamics. They are the concrete mathematical expressions of the energy conservation in nonextensive thermodynamics. Just as pointed out in Ref. [18], if the correlation (nonextensive) terms of whatever observable or interactions can be neglected, what is the origin of the nonextensivity of entropy? It is well-known that entropy should be a continuous function of the observables. For example, for a simple nonextensive system with $S = S(U,T)$, if the internal energy is extensive, is entropy possibly nonextensive? In addition, if Eq. (14) is true, one will lose Eqs. (10)[or (9)] and (4), which are crucial for the nonextensive theory. If Eq. (4) fails, we cannot, in fact, find even the entropy correlation given by Eq. (5). It is thus clear that like entropy, energy is also nonextensive in Tsallis statistics, while the standard energy additivity rule (14) is not suitable in Tsallis statistics because it directly violates the law of energy conservation.

Summing up, we have solved two long-standing problems in nonextensive statistical mechanics. The zeroth law of thermodynamics cannot be proved from theory, but it has been implicitly used in nonextensive statistical mechanics, while the so-called generalized zeroth law of thermodynamics derived by several authors may not be correct because it is derived on the basic of Eq. (8) and its corollary is obviously in contradiction with Eq. (8). The zeroth law of thermodynamics is a base of nonextensive statistical mechanics, while it is unnecessary to introduce the so-called generalized zeroth law of thermodynamics and the new concept of the physical temperature. The standard energy additivity rule (14) used widely by many researchers may not be suitable in nonextensive statistical mechanics because it violates the law of energy conservation and its corollary is obviously in contradiction with its premise. In nonextensive statistical mechanics, one has to use the pseudo-additivity energy rule which is consistent with the zeroth law of thermodynamics and satisfies the law of energy conservation. Finally, it is pointed out that the conclusions obtained here conform to Abe's standpoint[6], statistical mechanics may be modified but thermodynamics should remain unchanged.

**Acknowledgements**

This work was supported by the National Natural Science Foundation (No.10275051), People's Republic of China.